\documentclass[iop]{emulateapj}

\usepackage{natbib}
\usepackage{adjustbox}

\def\d{{\rm d}}

\newcommand{\beq}{\begin{equation}}
\newcommand{\eeq}{\end{equation}}

\begin{document}

\title{Clustering-based Redshift estimation: Comparison to Spectroscopic Redshifts} 

\shortauthors{Rahman et al.}

\author{
Mubdi Rahman\altaffilmark{1}, 
Brice M\'{e}nard\altaffilmark{1,2,3}, 
Ryan Scranton\altaffilmark{4},
Samuel J. Schmidt\altaffilmark{4}, 
Christopher B. Morrison\altaffilmark{5}
}

\altaffiltext{1}{Department of Physics and Astronomy, Johns Hopkins
  University, 3400 N. Charles Street, Baltimore, MD 21218}
\altaffiltext{2}{Kavli IPMU (WPI), the University of Tokyo, Kashiwa 277-8583, Japan}
\altaffiltext{3}{Alfred P. Sloan Fellow} 
\altaffiltext{4}{Department of Physics, University of California, One
  Shields Avenue, Davis, CA 95616, USA} 
\altaffiltext{5}{Areglander-Institut f\"{u}r Astronomie, Auf dem H\"{u}gel 71,   53121 Bonn, Germany}
\email{mubdi@pha.jhu.edu}

\submitted{Submitted to MNRAS}

\begin{abstract}

We investigate the potential and accuracy of clustering-based redshift estimation using the method proposed by \cite{menard13}. This technique enables the inference of redshift distributions from measurements of the spatial clustering of arbitrary sources, using a set of reference objects for which redshifts are known. We apply it to a sample of spectroscopic galaxies from the Sloan Digital Sky Survey and show that, after carefully controlling the sampling efficiency over the sky, we can estimate redshift distributions with high accuracy. Probing the full colour space of the SDSS galaxies, we show that we can recover the corresponding mean redshifts with an accuracy ranging from $\delta z=0.001$ to $0.01$. We indicate that this mapping can be used to infer the redshift probability distribution of a single galaxy. We show how the lack of information on the galaxy bias limits the accuracy of the inference and show comparisons between clustering redshifts and photometric redshifts for this dataset. This analysis demonstrates, using real data, that clustering-based redshift inference provides a powerful data-driven technique to explore the redshift distribution of arbitrary datasets, without any prior knowledge on the spectral energy distribution of the sources.

\end{abstract}


\section{Introduction}

Mapping celestial objects occurs fundamentally in two-dimensions. Exploring their physical properties, however, requires some knowledge of their distribution along the third dimension, distance. On extragalactic scales, distances are primarily estimated by combining measured redshifts with the expansion history of the Universe. The robustness of redshift estimation can vary dramatically depending on the type of source and the technique used. High quality spectroscopic redshifts are available only when one can detect and identify a high-contrast spectroscopic feature (emission/absorption lines or spectral break) at sufficient resolution, but such observations are typically expensive and restricted to bright objects. For the vast majority of extragalactic sources, distance estimates rely on photometric redshifts that require a priori knowledge of the type of object observed. They depend on models or spectroscopic training sets and are prone to catastrophic outliers due to degeneracies between the observed colours and redshift. As new surveys will soon map out several billion galaxies, the lack of robust distance estimates is becoming a serious limitation, hindering our exploration of the Universe.

Redshifts can, in principle, be inferred through a different approach: using information encoded within the clustering of matter, rather than the spectral energy distribution of sources. This idea has been discussed for more than thirty years, with attempts made based on angular cross-correlation measurements, such as those by \citet{seldner79} and \citet{phillipps87}. \citet{landy96} furthered the idea, demonstrating that a combination of auto- and cross-correlations between two populations of galaxies can be used to test whether the two samples overlap in redshift space. Along similar lines, \citet{ho08} used a combination of spatial auto- and cross-correlations with spectroscopic samples to constrain the first moments of the redshift distribution of the NVSS radio survey. Over the past eight years, several teams have designed methods aimed at characterizing redshift distributions from spatial correlations, including \citet{schneider06}, \citet{newman08}, \citet{matthews10} and \citet{mcquinn13}. They primarily considered future (LSST-like) surveys and focused on large-scale clustering (where the galaxy and dark matter fields are related through a linear bias) to determine redshift distributions with percent-level accuracy. This has been motivated by the requirements of upcoming photometric surveys designed to constrain the properties of dark energy (which are not met by the photometric redshift techniques currently available). While all of these methods are promising, their applicability to real datasets at the promised level of accuracy still needs to be demonstrated.

Another approach to clustering-based redshift inference has been proposed by \citet[][hereafter M13]{menard13}. By construction, their method does not aim at percent-level accuracy, but rather is designed to be directly applicable to existing datasets, optimizing the expected signal-to-noise by including small-scale clustering information (i.e. with $r<$ Mpc) to avoid systematic effects often affecting large-scale photometric calibration. Some demonstrative results were presented in \citet{schmidt13} with numerical simulations and in M13 with pilot studies using real datasets. In this paper, we implement the technique with a greater degree of sophistication (taking into account sampling considerations, cosmic variance, etc.) and test the reliability of our method by comparing clustering-based redshifts to spectroscopic redshift for galaxy populations selected from the Sloan Digital Sky Survey (SDSS). This work allows us to verify the expected accuracy of clustering-based redshift inference and demonstrate the potential of this new technique.

\section{Clustering-based redshift estimation}
\label{sec:clusterz}

Our approach is based on the method introduced by M13. We refer the reader to this paper for the detailed description of the formalism. In this section, we briefly re-introduce the main concepts.

We consider two populations of extragalactic objects: (i) a \emph{reference} population for which the angular positions and redshifts of each object are known. This population is characterized by a redshift distribution $\d{\rm N_r}/\d z$, a mean surface density $n_r$, and a total number of sources $N_r$; and (ii) an \emph{unknown} population for which angular positions are known but redshifts are not. Similarly, this population is characterized by the quantities $\d{\rm N_u}/\d z$, $n_u$ and $N_u$.

The basic principle is that, if the two populations do not overlap in redshift, their angular correlation is expected to be zero (ignoring gravitational lensing effects). As discussed by M13, in the ideal case of an unknown sample located within a narrow redshift range, one can accurately probe its redshift distribution by splitting the reference population into redshift slices $\delta z_i$ and measuring the angular or spatial correlations with the unknown population $w_{ur}(\theta,z_i)$ for each subsample $i$:
\begin{equation}
{\rm d N_u/\d} z\propto w_{ur}(\theta,z_i)\,.
\label{eq:propto}
\end{equation}
The spatial correlation $w_{ur}(\theta,z_i)$ is measured over some angular scale $\theta$ and is defined by
\begin{equation}
w_{ur}(\theta,z_i) = 
\frac{ \langle n_u(\theta,z_i) \rangle_r }{n_{u}} - 1\,,
\label{eq:estimator}
\end{equation}
where $\langle n_u(\theta,z_i) \rangle_r$ denotes the mean density estimate of the unknown sample around reference objects at redshift $z_i$. Following M13, we optimize the signal-to-noise (S/N) of our estimator by considering the integrated cross-correlation function
\begin{equation}
{\overline w_{ur}}(z) = \int_{\theta_{\rm min}}^{\theta_{\rm max}} \d\theta\, W(\theta)\, w_{ur}(\theta,z)
\label{eq:w_int}
\end{equation}
where $W(\theta)$ is a weight function, whose integral is normalized to unity, aimed at optimizing the overall S/N. To probe the same range of physical scales as a function of redshift, we set $(\theta_{min},\theta_{max})$ to match a fixed range of projected radii $(r_{p,min},r_{p,max})$ in physical space. Once a cross-correlation signal is found, the amplitude of the redshift distribution is simply obtained through the normalization
\begin{equation}
\int \d z\,\frac{\rm d N_u}{\d z} = {\rm N_u}\,.
\label{eq:normalization}
\end{equation}
The normalization is dependent on all sources in the unknown sample existing within the redshift range of the reference sample.

As in M13, departing from this ideal situation will only cause a modest loss of accuracy in many cases, and redshift inference can still be made with sufficient precision for a large range of astrophysical applications. The degree of departure from an exact solution can be estimated by examining the relative contributions of the terms contributing to the angular correlation function: the redshift distribution ($\d{\rm N_u}/\d z$) and the bias-related clustering amplitude of each population. If, over the redshift range $\Delta z$, the relative variation of $\d{\rm N_u}/\d z$ dominates over that of $b_u(z)$, i.e.
\begin{equation}
\frac{\d \log {\rm \d N_u/\d}z}{\d z}
\gg
\frac{\d \log {\overline b_u}}{\d z}
\label{eq:recovery_criteria}
\end{equation}
we approach the case where $\d{\rm N_u}/\d z\rightarrow {\rm N_u}\,\delta_D(z-z_0)$, and Eq.~\ref{eq:propto} and \ref{eq:normalization} can be used to to infer $\d{\rm N_u}/\d z$. However, this inference is only valid up to a finite accuracy. 

The main limitation of the technique is the absence of clustering amplitude information, which places a limitation on its accuracy. With our current approach, we note that we only require the derivative of these clustering amplitudes with redshift, i.e. $\d{\rm {\overline b_r}}/\d z$ and $\d{\rm {\overline b_u}}/\d z$. In this analysis we will infer $\d{\rm {\overline b_r}}/\d z$ from the measured auto-correlation of the reference population as a function of redshift, using the same range of scales and weighting as used in Eq.~\ref{eq:w_int}. The determination of this quantity is presented in the Appendix~\ref{sec:refclustering}.
We will then treat the clustering amplitude of the unknown population in two ways:
\begin{itemize}
\item Firstly, we will neglect its contribution, i.e. we assume $\d{\rm {\overline b_u}}/\d z=0$.
\item Secondly, we assume that, on average, the clustering amplitude integrated over the range of scales considered evolves linearly with redshift: $\d{\rm {\overline b_u}}/\d z=1$. This is suggested by the redshift evolution of the amplitude of the galaxy auto-correlation function (see Appendix~\ref{sec:refclustering} for more detail).
\end{itemize}
Comparing the two estimators will enable an estimation of the uncertainty caused by the absence of clustering amplitude information for the unkown population. According to M13, even ignoring the evolution of the clustering amplitude of the unknown sample (i.e. using $\d{\rm {\overline b_u}}/\d z=0$), the technique is expected to provide a estimate of mean redshifts with accuracy  $\delta z \sim 0.01$ for $z<1$ galaxy populations. We will test this below using a spectroscopic dataset for which redshifts are accurately known.


\begin{figure}
	\begin{center}
	\includegraphics[scale=0.47]{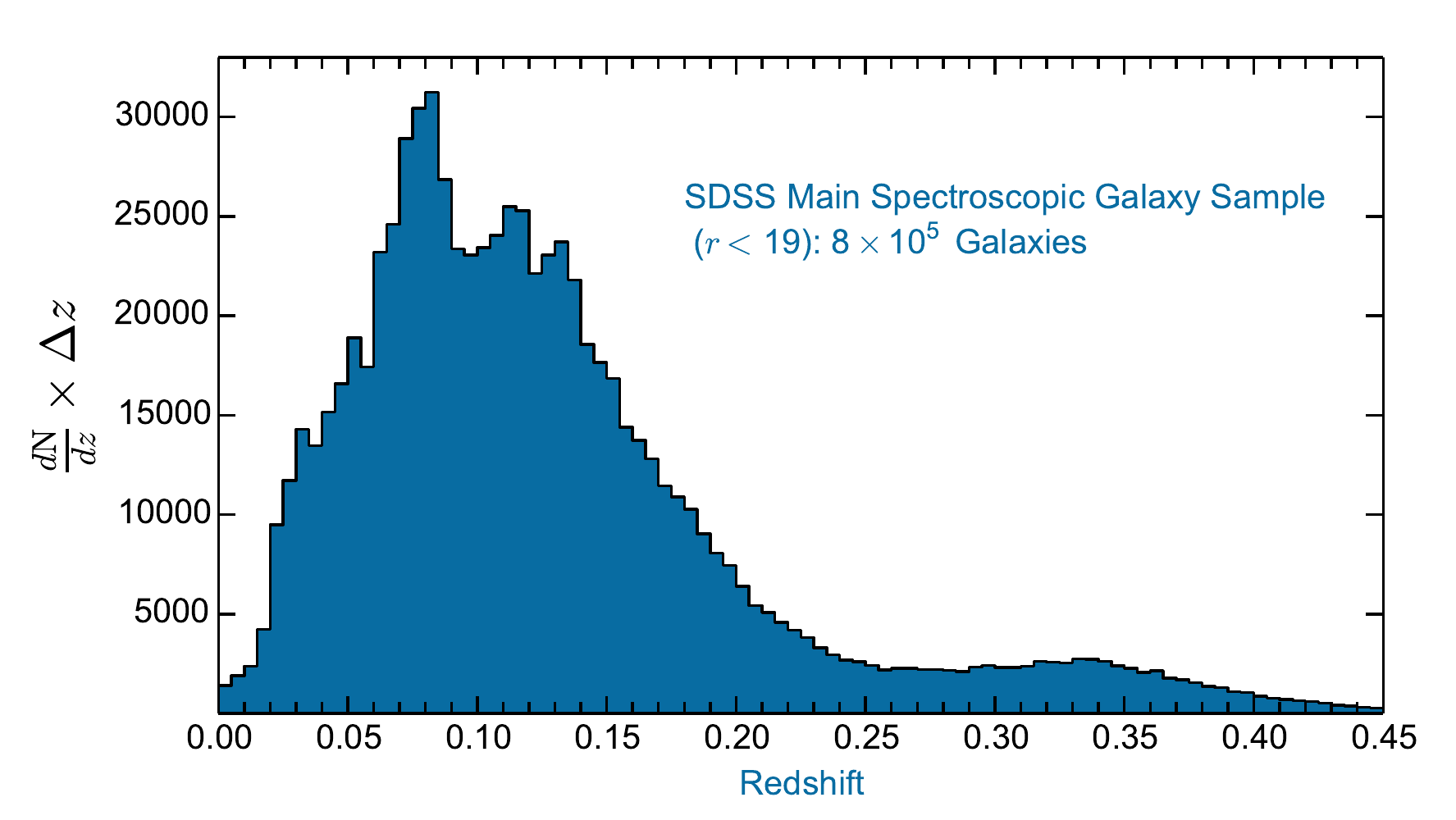}
	\end{center}
		\caption{The redshift distribution of the SDSS legacy spectroscopic galaxy
		sample with $r_{model} < 19$, presented with bin width $\Delta z = 0.005$.}
		 \label{fig:galdist}
\end{figure}



\begin{figure*}[t]
	\begin{center}
	\includegraphics[scale=0.7]{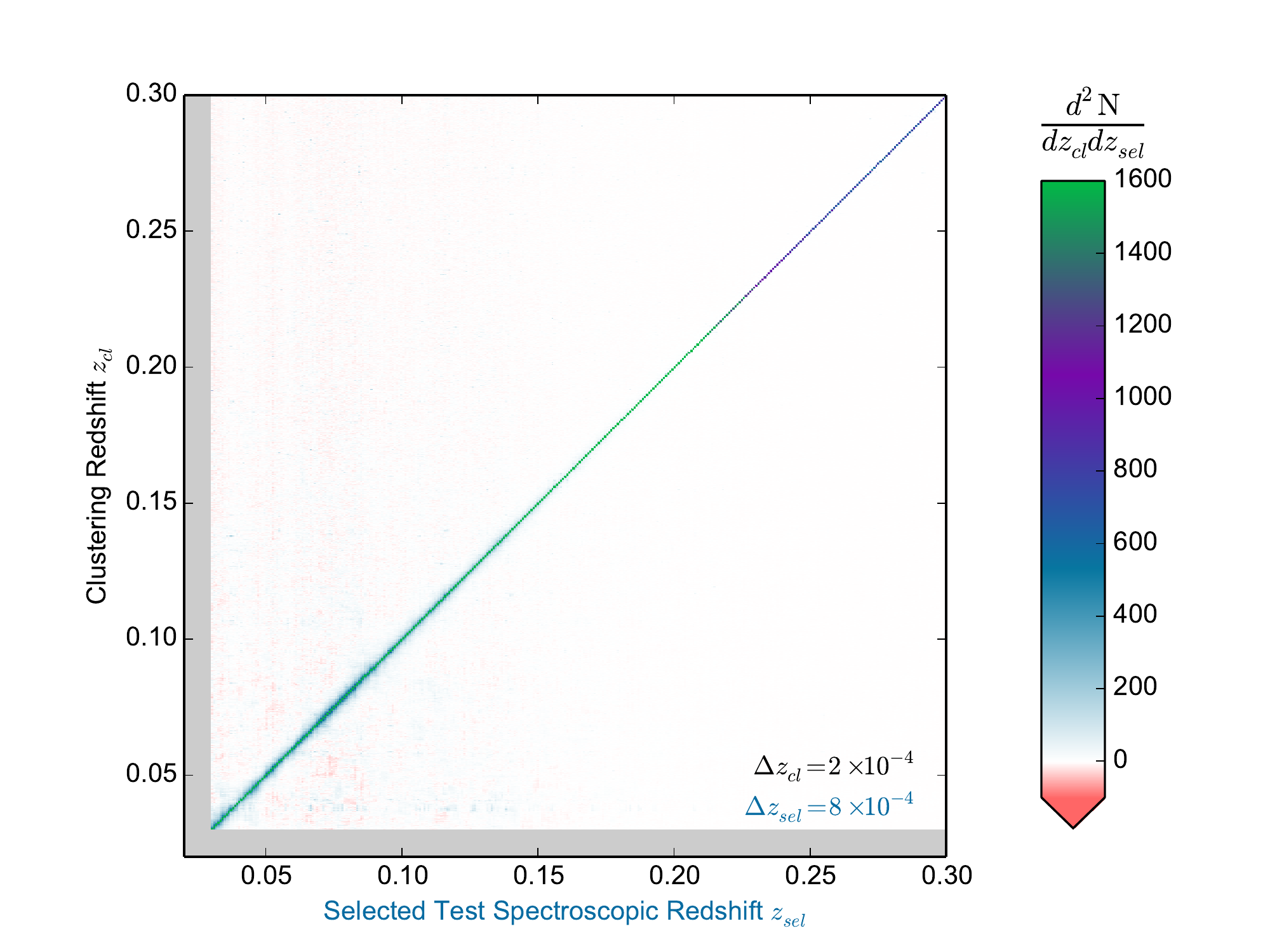}
	\end{center}
		\caption{The comparison between clustering redshifts $z_{cl}$ and spectroscopic redshifts $z_{sel}$ for samples selected in narrow redshift bins. The figure shows the density distribution $\d^2{\rm N}/\d z_{cl}\d z_{sel}$ sampled with $\Delta z_{cl}=2\times10^{-4}$ and $\Delta z_{sel}=8\times10^{-4}$, derived from about half a million cross-correlation measurements over the Northern Galactic Cap of the SDSS. A given column corresponds to the redshift distribution inferred for galaxies selected with a given spectroscopic redshift $z_{\rm sel}$.
}
\label{fig:mpagalheatmap}
~\vspace{.3cm}
\end{figure*}


\section{Data Analysis}

\subsection{The Dataset}

We use the Legacy Spectroscopic Galaxy Sample from the Sloan Digital Sky Survey \citep[SDSS;][]{york00,sdssdr7}. This sample consists of nearly one million galaxies distributed over $>8000$ square degrees. It is composed of two selections: the ``Main Galaxy'' sample \citep{strauss02}, a complete sample selected with $r_{\rm p}<17.77\,$mag, and the Luminous Red Galaxy (LRG) sample \citep{eisenstein01}, extending to fainter magnitudes. To ensure on-sky uniformity of the sample, we use a limiting magnitude $r_{\rm model}=19$ mag, restricting the entire sample to 791,546 galaxies. The corresponding redshift distribution is shown in Figure \ref{fig:galdist}. Typical spectroscopic redshift errors for these sources are of order $10^{-4}$. 

\subsection{Density Estimation}
\label{sec:density_estimation}

The ability to infer robust redshift distributions from the clustering-based technique relies on accurate source density estimation. This requires measuring densities within well-defined angular apertures, taking into account regions of unreliable photometric data coming from missing data, poor photometry, poor sky background estimation, and artifacts from bright stars, satellite tracks, etc. We handle this complexity and perform calculations on the celestial sphere with the astro-STOMP library\footnote{The open-source STOMP library is available at \url{https://code.google.com/p/astro-stomp/}}.

We minimize potential biases from Galactic dust extinction, as well as keep the overall footprint simple, by limiting the present analysis to a region in the Northern Galactic Cap, bounded by $100\degr < \alpha < 280\degr$ and $-11\degr < \delta < 80\degr$. This field covers about 5400 square degrees. We measure spatial cross-correlations within the range of physical scales $300\;{\rm kpc}<r_p<3\;{\rm Mpc}$ at the redshift of each reference subsample. This corresponds to an inner radius of 67\arcsec at $z = 0.3$ and ensures that the fiber collision exclusion (55\arcsec) is avoided in the measurement. 
As with M13, we choose a weight function $W(\theta) \propto \theta^{-0.8}$ (see Eq.~\ref{eq:w_int}), which is expected to mimic the spatial dependence of the galaxy correlation function. In order to reach the level of precision required in our analysis, we need to account for the small loss of area induced by the presence of additional fibers within the angular apertures (of order a few percent) to obtain sufficiently accurate density estimates. We compensate for this by making the correction to the mean area estimate. 

We begin by exploring the ideal regime for clustering-based redshift inference: selecting populations of galaxies in a narrow redshift bin limit. As discussed in Sec.~\ref{sec:clusterz}, this regime is expected to provide nearly exact results. 
We use galaxies between $0.03 < z < 0.3$ and define a set of ``selected'' subsamples with $\Delta z_{\rm sel} = 8 \times 10^{-4}$. Similarly, we define a sequence of reference subsamples with $\Delta z_{ref} = 2\times 10^{-4}$, amounting to 1400 redshift bins. In this narrow bin limit, we can use clustering measurements to locate the test samples in redshift space. This can be done without any additional information of the clustering amplitude of each population. Consequently, we use ${\rm \d {\overline b_u}/\d} z$=0. For each test sample, we measure the weighted, integrated cross-correlation with the  reference subsamples (Eq.~\ref{eq:w_int}) and estimate the normalized redshift distribution using Equation \ref{eq:normalization}.

We present the corresponding set of measurements in Figure~\ref{fig:mpagalheatmap}, showing the estimated density of selected galaxies, $\d^2{\rm N}/\d z_{cl}\d z_{sel}$, as a function of clustering redshift. Each column of this figure shows the estimated redshift distribution for a population of spectroscopic objects selected within a redshift bin $\Delta z = 8 \times 10^{-4}$. Overall, we find a good agreement between the redshift of each selected population and the clustering-based redshift estimate as indicated by the $z_{cl}=z_{sel}$ line. Away from this line, our technique does not indicate any problematic signal at a level greater than a few percent. The width of the signal in the vertical direction appears to increase toward lower redshifts. This effect, due to the redshift evolution of the clustering amplitude of the selected objects, defines the response function of our technique, i.e. the redshift distribution observed for an input sample located at a single redshift. This quantity is characterized and discussed in more detail in Appendix~\ref{sec:response}. The accuracy of the estimate of the peak position is actually higher than the size of each redshift bin. More than 90\% of the signal in these distributions is located within a few resolution elements around the centre of the distribution, with $\Delta z_r < 10^{-3}$.


\begin{figure*}
	\begin{center}
	\includegraphics[scale=0.7]{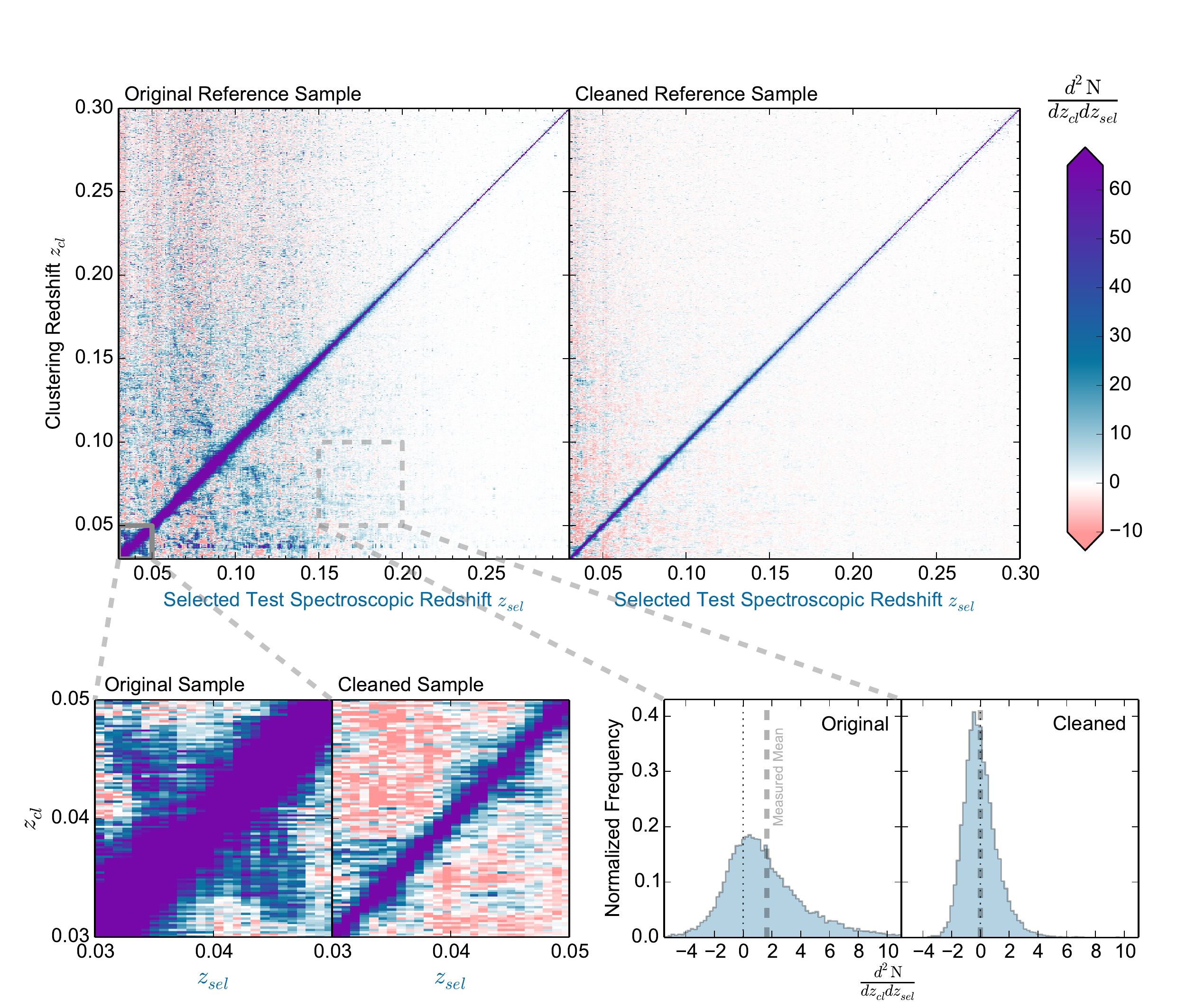}
	\end{center}
		\caption{\textit{Top:} the density distribution $\d^2{\rm N}/\d z_{cl}\d z_{sel}$ as shown in Figure \ref{fig:mpagalheatmap} but showing only the first few percent of its dynamical range ($\d^2{\rm N}/\d z_{cl}\d z_{sel}<60$). The intrinsic clustering of galaxies from the reference sample imprints correlated patterns which can contaminate clustering-redshift estimation, as shown in the left panel. A spatial homogenization of the density distribution of the reference sample can be applied to suppress these spurious effects, as shown in the right panel. \textit{Bottom-left:} A zoom-in at the redshift range where the Hercules Supercluster is located. The peculiar velocities of galaxies within the cluster give rise to a signal perpendicular to the diagonal. This effect is almost entirely removed when using the cleaned reference sample.	\textit{Bottom-right:} A histogram comparing the distribution of redshift densities where no signal is expected. The cleaned reference sample allows a reduction in the scatter as well as a more robust zero point estimate as indicated by the dashed line.
		}
		\label{fig:cleancompare}
		~\vspace{.3cm}
\end{figure*}


\subsection{Noise Properties}

To highlight low-level features in Figure~\ref{fig:mpagalheatmap}, we present a clipped version of the distribution in Figure \ref{fig:cleancompare}, limiting the maximum value of $\d^2{\rm N}/\d z_{cl}\d z_{sel}$ to 60, compared to about $1,600$ in the previous version. Consequently, these density maps show fluctuations at the few percent level. The figure demonstrates that, away from the $z_{cl}=z_{sel}$ track, the statistically estimated $\d{\rm N}/\d z$ oscillates around zero with a level in agreement with the expected Poisson noise. While the typical source density of each selected sample is less than one object per square degree, the angular aperture used in the cross-correlation measurements covers about 0.1 square degree at $z=0.3$. In regions where we expect no clustering signal (such as away from $z_{cl}=z_{sel}$) the mean angular correlation, and equivalently the measured overdensity, is expected to be consistent with zero. As the resolution of the density map increases, our sampling of the underlying Poisson distribution becomes finer and, given its asymmetry, we expect most of the pixels to display values that are slightly below the overall average. This gives rise to the faint red background in the figure. However, averaging over larger groups of pixels leads to values consistent with zero. This effect becomes stronger when the sample size decreases or when the aperture becomes smaller; the probability of having no object falling within the selected aperture is higher. 

Towards the lower left corner of the figure, structure appears away from the $z_{cl} = z_{sel}$ track, significantly above the noise level. There are due to three types of effects:
\begin{enumerate}
\item The presence of a massive galaxy cluster in the reference sample leads to a substantial increase in the sampling of the corresponding region of the sky. This creates a spurious signal each time a selected sample presents an overdensity in the same region of the sky. This gives rise to a series of correlations distributed horizontally, located at the redshifts of the massive clusters in the reference sample. This effect can be seen in Figure \ref{fig:cleancompare}: a horizontal feature is present at $z_{cl} = 0.037$. It extends from $z_{\rm sel}\simeq0.05$ to about 0.2. This is caused by the presence of the Hercules Supercluster located at this redshift, the largest cluster in the local universe \citep[$M \sim 10^{16}\, \textrm{M}_\odot$;][]{barmby98}. 
\item The radial velocity of an extragalactic object has contributions from both the Hubble flow and its own peculiar velocity. Since only cosmological redshifts correlate with distance, regimes in which peculiar velocities are high will create spurious signals in the $\d^2{\rm N}/\d z_{cl}\d z_{sel}$ estimates. This effect is strongest when massive galaxy clusters are present. For galaxy clusters with a gravitational potential of order $\Delta v =10^3\,{\rm km\,s^{-1}}$, this amounts to $\Delta z\sim 10^{-3}$. The contribution of peculiar velocities affects our ability to properly infer redshifts solely due to the Hubble flow. We note that this degeneracy cannot be removed based on velocity information alone. This leads to a spurious correlation signal perpendicular to and symmetric about the $z_{cl} = z_{sel}$ line. Objects moving around the cluster with a negative line-of-sight velocity, inferred to have a higher redshift, will correlate with cluster objects with a positive velocity, inferred to have a lower redshift. This can be seen in the region $z_{sel}\sim z_{cl}\sim 0.037$. The peculiar velocity effect of the cluster is illustrated as a perpendicular spread of correlation signal (bottom-left of Figure \ref{fig:cleancompare}). The velocity spread of the signal ($\sim 2000$ km s$^{-1}$) is consistent with the mass of the supercluster.
    \item Chance superpositions of large scale structure from two different redshifts, when projected onto the sky, produce a artificial correlation signal: if two structures well-separated in redshift overlap on the sky, the reference galaxies at one redshift will measure an overdensity in the selected sample of the second redshift. Similarly, the reference galaxies at the second redshift will measure an overdensity in the first redshift. Consequently, these spurious correlations appear symmetric about the $z_{cl} = z_{sel}$ line. An example of the first effect is seen in Figure \ref{fig:cleancompare} as structure at $(z_{sel}, z_{cl}) = (0.08, 0.11)$ and symmetrically at $(z_{sel}, z_{cl}) = (0.11, 0.08)$. 
\end{enumerate}
These three effects explain the origin of virtually all the structures appearing in the left panel of Figure~\ref{fig:cleancompare}. The larger volumes sampled at higher redshift minimize the effect of cosmic variance, thereby decreasing the amplitude of these artificial signals. Since the origin of the spurious correlations is primarily due to spatial inhomogeneity of the reference population, we can filter the reference sample to minimize these effects. We describe this procedure below.

\subsection{Cleaning the Reference Sample}
\label{sec:cleaning}

The spurious correlations described above are mainly due to the inhomogeneous sampling arising from the clustered distribution of objects in the reference sample. The issues from clustering can be addressed by homogenizing its spatial distribution. This is done by optimally weighting each reference galaxy based on its local density and propagating these weights when characterizing the spatial correlations. 
Applying this procedure in angular space would maximize the number of usable galaxies in the reference sample but it would not remove the spurious correlation due to the peculiar velocity effects. As the main goal of the present analysis is not to optimize the statistical power of clustering-based redshift inference but simply test its accuracy, we homogenize the sample through a simple selection of the data, keeping only regions of the sky for which the galaxy density does not strongly depart from its mean value and applying an appropriate masking to ensure that effects due to peculiar velocities are neglegible.
%
To do so, we split the reference sample into 60 equal-area regions on-sky, each covering  $\sim$90 square degrees, and select redshift bins with $\Delta z = 10^{-3}$ in the range $0.03 < z < 0.45$, corresponding to 420 sequential bins. We measure the densities of the cells and keep only those for which the value is within $2\,\sigma$ of the mean. For each region exceeding it, we remove all galaxies from the reference sample within $\Delta z = \pm 1.5 \times 10^{-3}$. We refer to the remaining galaxies as the \emph{cleaned reference sample}. The fraction of regions masked is about 30\% at $z=0.03$ and decreases to less than 5\% at $z=0.3$. Since these regions are selected based on the excess number of galaxies within them, a greater fraction of the galaxies will be masked than the fraction of area spanned on sky. Consequently, this aggressive homogenization strategy masks about 45\% of the galaxies, leaving a final cleaned reference sample of 442,000 objects. While this procedure removes a fair fraction of the reference galaxies, the cleaned reference sample is robust to all three spurious correlation effects mentioned above, thereby decreasing the overall noise level in spite of the substantially decreased reference sample size.

The right side of Figure~\ref{fig:cleancompare} shows the resulting clipped redshift distribution estimates when using the cleaned reference sample defined above. As compared to the original density plot, the vast majority of the spurious signal disappears. The overdensity originating from the Hercules Supercluster has been suppressed without any a priori knowledge of its location. This is highlighted in the insets shown on the bottom left. The related horizontal track described above, as well as almost all the large-scale overdensities, are no longer detectable. The overall $\d^2{\rm N}/\d z_{cl}\d z_{sel}$ signal is also more concentrated onto the $z_{cl}=z_{sel}$ line. This leads to a reduction in the overall noise level, despite the use of fewer reference galaxies from the aggressive masking strategy. This can be seen in the subpanel shown in the bottom right of the figure, presenting the distribution of $\d^2{\rm N}/\d z_{cl}\d z_{sel}$ values. The use of the cleaned reference sample leads to better control of the zero point, i.e. the estimate of the mean galaxy density. In addition, we observe a significant reduction in the width of the distribution, corresponding to a noise level below 0.3\%. We note that the cleaning process is only applied to the reference sample. No operation is applied to the set of selected test galaxy subsamples.

This cleaned reference sample can be used to infer the redshift distribution of any unknown population spanning the redshift range $0.03 < z < 0.45$. It will be used in a companion paper (Rahman et al., in prep) to investigate the distribution of clustering redshifts for the entire photometric sample of SDSS galaxies. This analysis demonstrates our ability to robustly estimate redshift for populations selected within a narrow redshift range -- the regime in which the method is expected to provide us with nearly exact results.


\begin{figure*}
	\begin{center}
		\includegraphics[scale=.7]{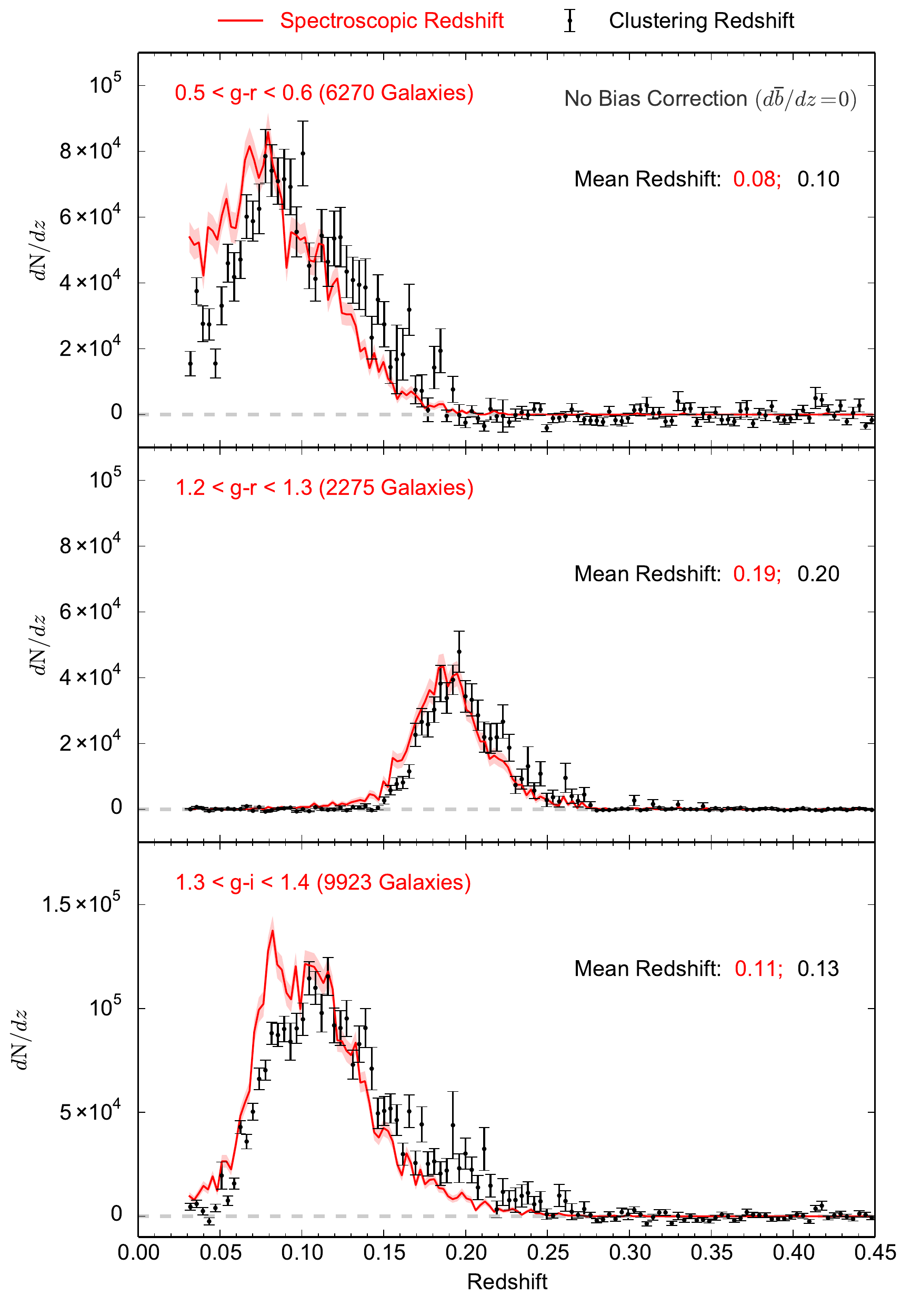}
	\end{center}
    \vspace{-0.3cm}
\caption{A comparison of the spectroscopic (red line) and clustering redshift distribution (black points) for galaxies selected with limiting magnitude $r < 17.77$ and three different colour cuts: $0.5 < g-r < 0.6$ (top), $1.2 < g-r < 1.3$ (middle), and $1.3 < g-i < 1.4$ (bottom). Here we ignore the possible redshift evolution of the clustering amplitude of the galaxies (i.e. we use $d\overline{b}_{sel}/dz = 0$). The mean clustering-redshifts agree with the spectroscopic ones within $\delta z=0.02$.
\label{fig:specialsamples_raw}}
~\vspace{.3cm}
\end{figure*}



\begin{figure*}
	\begin{center}
    	\includegraphics[scale=.7]{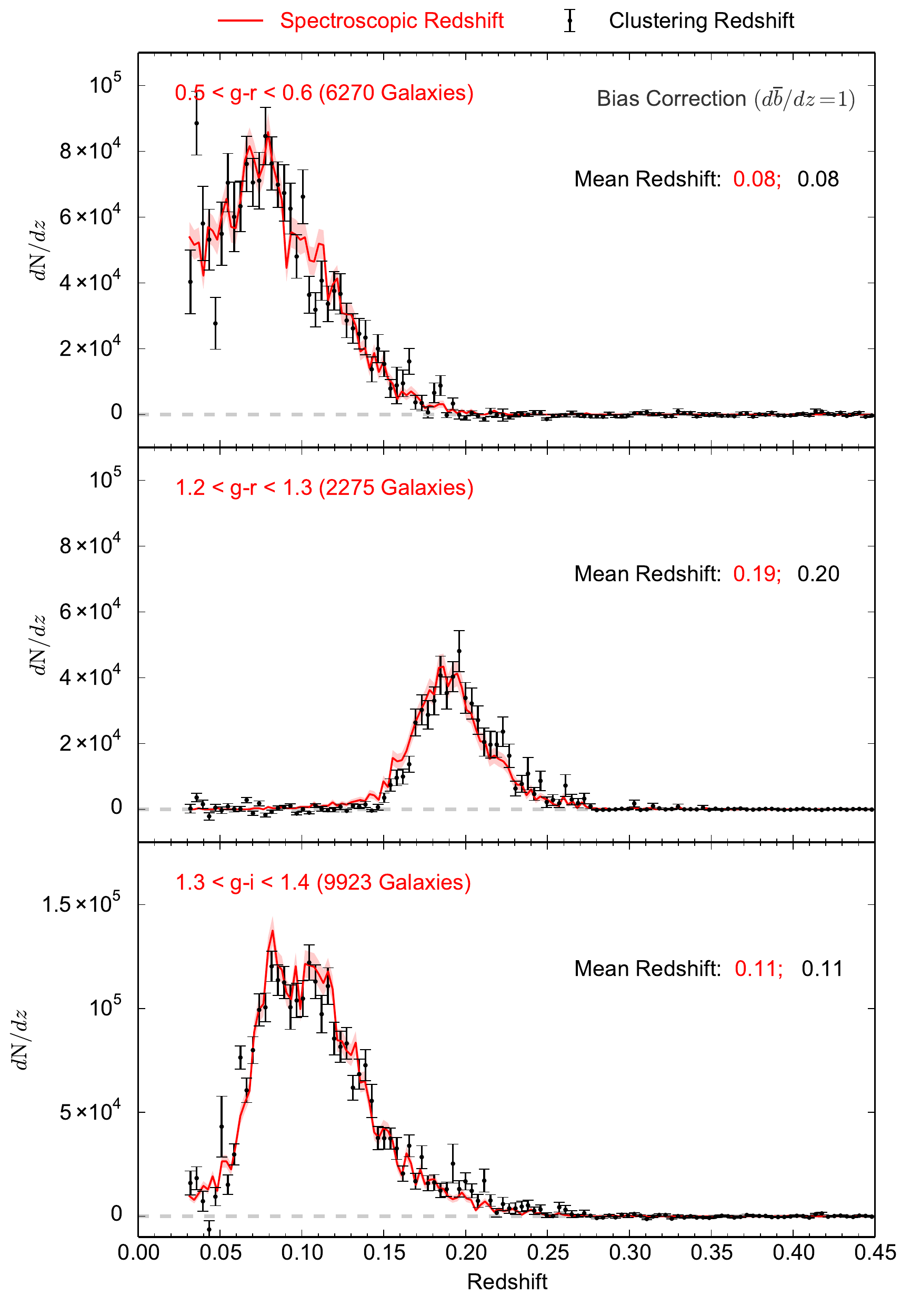}
        \vspace{-0.3cm}
    	\caption{Same as Figure \ref{fig:specialsamples_raw}, but using a linear redshift evolution correction, i.e. ${\rm d\overline{b}/d}z = 1$. The mean clustering-redshifts agree with the spectroscopic ones within a $\delta z$ better than $0.01$.
    	}
	    \label{fig:specialsamples_lin}
	\end{center}
~\vspace{.3cm}
\end{figure*}



\begin{figure*}[th]
	\begin{center}
		\hspace*{-1cm}\includegraphics[scale=.65]{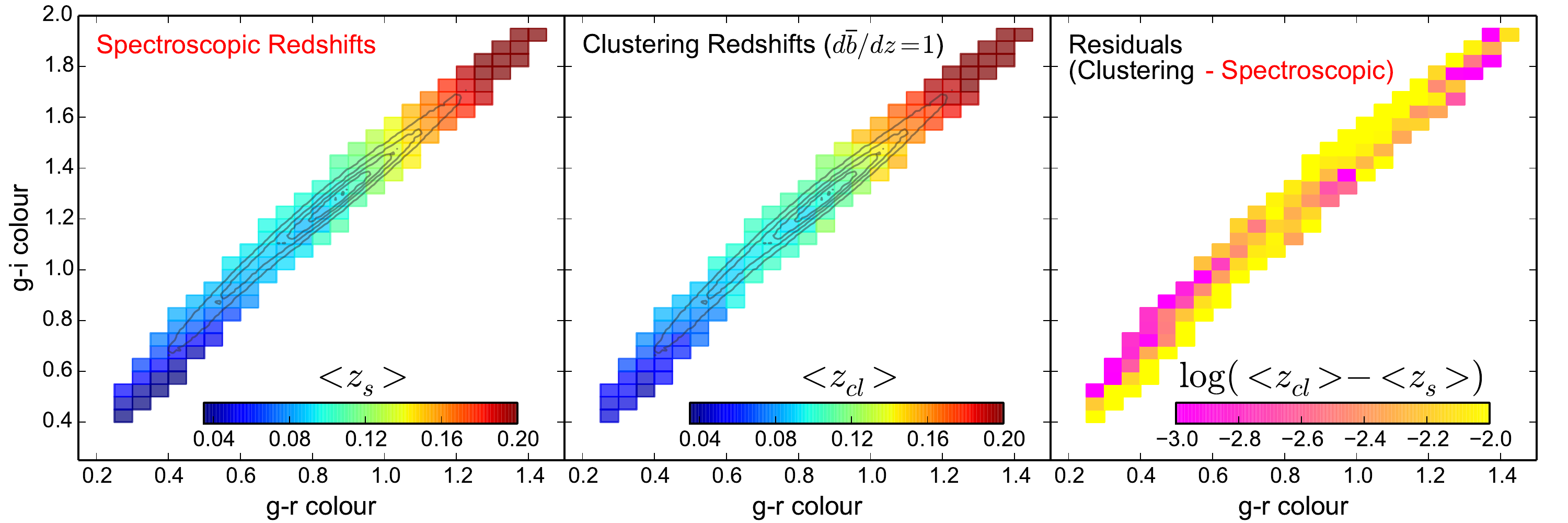}
	\end{center}
	\caption{The distribution of the mean redshifts of SDSS spectrosopic galaxies selected as a function of their $g-r$ and $g-i$ colours using  spectroscopic redshifts (left) and clustering redshifts with ${\rm d\overline{b}/d}z = 1$ (centre). The solid lines represent the galaxy number count density. We have limited our mapping to colour cells with more 1000 sources. The right panel shows the difference between the two distributions and shows that clustering redshifts are slightly overestimated by $\delta z$ ranging from 0.001 to 0.01.}
	\label{fig:meanredshift_col}
~\vspace{.3cm}
\end{figure*}


\section{Results with Photometric Samples}
\label{sec:results}

Having demonstrated the ability to estimate the distributions for narrow redshift samples, we now apply our method to a more generic scenario: galaxies photometrically selected, thus expected to span a finite redshift range. As an illustration, we consider three arbitrary galaxy samples selected by their photometric properties. We use a limiting magnitude of $r < 17.77$ and three colour cuts: 
\begin{eqnarray}
0.5 < g-r < 0.6 & ~~(6270~{\rm galaxies})\nonumber\\
1.3 < g-i < 1.4 & ~~(2275~{\rm galaxies})\nonumber\\
1.2 < g-r < 1.3 & ~~(9923~{\rm galaxies})
\end{eqnarray}
The corresponding densities on the sky are 0.5-2 sources/deg$^2$. Figure~\ref{fig:specialsamples_raw} shows the spectroscopic redshift distribution of these galaxies with the red line. The red contours indicate the estimated Poisson noise of each redshift bin, determined through the variance of the measurement. The selected samples have mean spectroscopic redshifts of 0.08, 0.19 and 0.11, respectively. The width of the distributions is of order $\Delta z=0.05$. We compare these ``true'' redshift distributions to those inferred through clustering redshifts using the cleaned reference sample, as defined in section~\ref{sec:cleaning}. Further, to fully simulate a practical application to real datasets and work with two distinct samples, we exclude the selected galaxies from our cleaned reference sample. We use the parameters listed in section~\ref{sec:density_estimation} and an estimation of the redshift dependence of the reference sample clustering amplitude $\d{\rm {\overline b_r}}/\d z$ measured from its auto-correlation (see Appendix \ref{sec:refclustering} for more detail).

We first consider the simplest case for which we ignore the redshift evolution of the clustering amplitude of the unknown population, i.e. we use $\d{\rm {\overline b_u}}/\d z=0$. The results are shown with the black data points. We estimate the errors from the variance of the mean density measurement. Overall, we observe some level of agreement. In all three examples, the mean clustering redshifts match the spectroscopic ones within $\Delta z = 0.02$. We note that this is in agreement with the predicted error level from M13. We also note that, even if the redshift evolution of the galaxy bias is unknonw, a useful property of the clustering redshift technique is the ability to check for the absence of galaxies in a given redshift range. We note that without any assumption, this method enables the inference of the redshift interval over which the photometric sample is distributed, without any assumption on the nature or the spectral energy distribution of the sources.

We now assume that the redshift evolution of the bias of the unknown population is characterized by $\d{\rm {\overline b_u}}/\d z=1$, i.e. it evolves linearly with redshift. While this is not expected to be exact, it appears to provide a first approximation to the redshift dependence of galaxies selected by stellar mass at $z\lesssim1$ (see Appendix~\ref{sec:refclustering}). The corresponding results are shown in Figure~\ref{fig:specialsamples_lin}. The clustering redshift distributions obtained with this simple assumption are found to be in excellent agreement with the spectroscopic redshifts. For each sample considered, the mean clustering redshifts match the spectroscopic one within $\Delta z=0.01$. The shape of the redshift distributions also appear to be in good agreement with the spectrosopic redshift distributions. We can even observe the bimodality in the distribution of the sample selected with $1.2 < g-r < 1.3$.


\subsection{Generalization to the Full Colour Space}

Having illustrated the potential of the clustering-redshift technique with three arbitrary colour-selected samples, we generalize our analysis to the entire colour space spanned by the SDSS legacy spectroscopic galaxies. To do so, we select galaxies as a function of their $g-r$ and $g-i$ colours. We divide this colour space into square cells with width $\Delta {\rm colour}=0.05\,$mag, and for each of them we measure the \emph{mean} spectroscopic redshift. We limit this analysis to cells containing a minimum of 1000 sources. The results are shown in the left panel of Figure~\ref{fig:meanredshift_col}. The solid contours indicate the number density of galaxies. Most of the SDSS spectroscopic galaxies have colours $g-r\sim0.9$ and $g-i\sim1.2$ mag. The colour scale indicates the mean redshift in each galaxy color cell. As expected, higher redshift galaxies appear redder. The distribution indicates that the $g-r$ and $g-i$ colours are highly correlated with each other and with mean redshift. One can however observe that, at a fixed $g-r$ colour, galaxies redder in $g-i$ appear to be at higher redshift.

We apply the clustering redshift technique to the same sample, using the same colour sampling. We do so using the linear bias correction as described above ($\d{\rm {\overline b_u}}/\d z=1$). The results are presented in the middle panel of the figure. As can be seen, the agreement between the mean spectroscopic redshifts and clustering redshifts is remarkable over the entire colour space. The differences between the two panels are nearly indistinguishable, demonstrating the generalization of the results found in the previous section. In order to reveal these differences, we show the residuals in the right panel, i.e. $\langle z_s\rangle - \langle z_{cl}\rangle$. These residuals are all positive, showing that the clustering redshifts tend to be slightly lower than the spectroscopic determination. This reflects that fact that $\d{\rm {\overline b_u}}/\d z=1$ corresponds to a redshift dependence that is slightly too shallow. Nevertheless, the amplitude of the residuals indicate that the difference in the mean redshift between the two estimators is smaller than $10^{-2}$, reaching $10^{-3}$ in some regions of the colour space.

This application to the SDSS legacy spectroscopic galaxies demonstrates the potential of the data-driven technique of clustering-based redshift estimation. We note that the accuracy we found is in agreement with the theoretical expectations presented in M13. For samples with a redshift distributions characterized by $\Delta z\sim0.05$, assuming $\d{\rm {\overline b_u}}/\d z=0$ or 1, we expect mean clustering redshifts to be accurate within 0.02. Our application to the entire colour space of SDSS spectroscopic galaxies has demonstrated this property.

\subsection{Inferring the Redshift PDF of a Single Galaxy}

Redshift estimation based on photometric information can be described as the mapping that connects volume elements in the space of photometric observables to redshift space. Photometric redshifts determine this mapping with a calibration based on theoretical or observed sets of spectral energy distributions. Our clustering-based estimation aims at determining the same mapping but using spatial correlations.

Figure~\ref{fig:meanredshift_col} is an example of such a mapping, connecting galaxy colour space to redshift space. In each cell of colour space, the application of the clustering-redshift technique produces an estimate of the corresponding redshift distribution. The mapping of the entire space can be used to infer the redshift probability distribution function (PDF) of a single galaxy, either by using the redshift distribution of the corresponding colour cell, or by using more advanced interpolation techniques and making use of information including nearby cells to increase the accuracy of the estimate. This is analogous to the process used to define the redshift PDF of a single galaxy with classical photometric redshifts: using the comparison between observed colours and the spectral energy distributions of modelled or observed galaxies. With clustering redshifts, it is determined using a set of spatial correlation functions.

There exists a unique mapping between a given space of photometric observables and redshift space. Every photometric cell $j$ maps onto a redshift distribution of finite extent $\Delta z_j$. Certain regions of this space may map onto multimodal regions of redshift space due to \emph{intrinsic} degeneracies in the mapping itself. There is a limit to how much redshift information can be extracted from the photometry and it applies to both photometric redshifts and clustering-based redshifts in the same way. In the case of a photometric cell mapping onto a multimodal redshift distribution, if subsampling in the other photometric dimensions does not break the redshift degeneracy, all the information used to map the photometric space onto the redshift space has been exhausted.

\subsection{Comparison to Photometric Redshifts}


\begin{figure*}[t]
\begin{center}
	\includegraphics[scale=0.7]{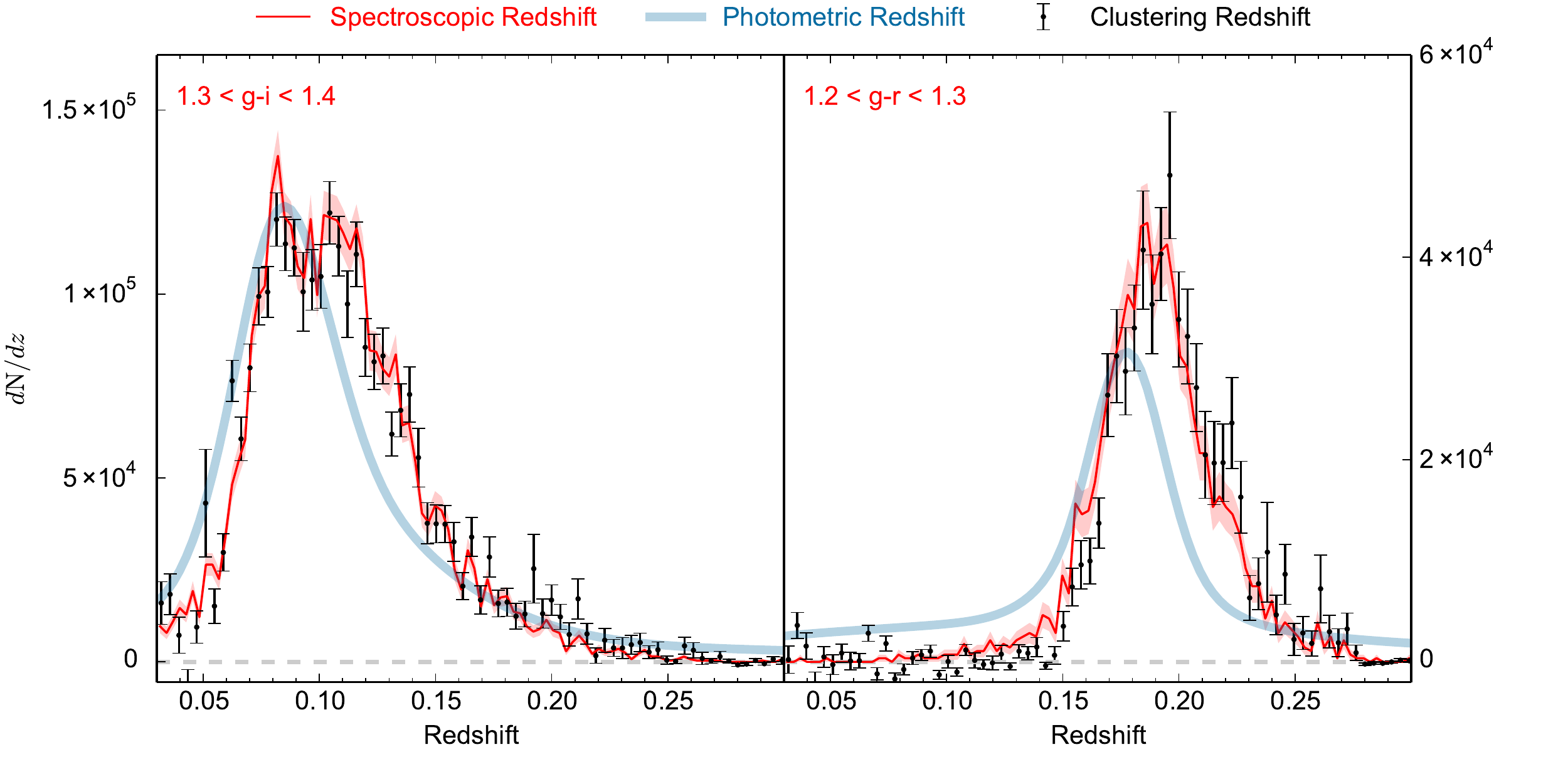}
\caption{A comparison of the spectroscopic (red), clustering (black) and photometric (blue) redshifts for two colour cut samples from Figure \ref{fig:specialsamples_raw}. The photometric redshifts are taken from the KD-Tree nearest neighbour technique \citep{csabai07} using the full probability distributions for each source. The accuracy of the clustering redshift distribution is greater than that from photometric redshifts. We point out the ability for clustering-redshifts to estimate small-scale structure in the distribution, such as the bimodality seen in the left panel $-$ and not captured by photometric redshifts.}
\label{fig:photoz}
\end{center}
~\vspace{.2cm}
\end{figure*}


To compare our clustering redshifts with photometric redshifts, we use the KD-Tree nearest neighbour redshifts from
\citet[][]{csabai07}, determining their the total redshift probability distribution. The individual probability distribution of a given galaxy was modeled as gaussian using the mean redshift and quoted uncertainty, producing the total redshift distribution by summing over all galaxies in the sample.
In Figure \ref{fig:photoz} we show the spectroscopic, photometric and clustering redshift estimates for two of the three galaxy samples introduced above, i.e. $1.3 < g-i < 1.4$ and $1.2 < g-r < 1.3$. We find that the mean redshift of the three estimators is, in those two cases, relatively similar. However, the overall shape of the redshift distribution indicated by the photometric redshifts significantly departs from the spectroscopic distribution. The peaks of the photometric distributions do not match those of the two other estimators. We note that in each case, the photometric redshift distribution displays a high redshift tail extending beyond the maximum redshift shown in the figure. For the sample selected with $1.2 < g-r < 1.3$, photometric redshifts indicate the presence of a substantial amount of galaxies at $z<0.12$ while the two other estimates indicate virtually no objects in that range. The bimodality of the distribution shown in both the spectroscopic and clustering redshift distributions for the sample selected with $1.3 < g-i < 1.4$ is not reproduced by the photometric redshifts. This comparison highlights the ability of the clustering-based redshifts to determine the existence of a population at any given redshift, and its robustness to catastrophic failures that plague photometric redshifts.

\section{Conclusions}

We have investigated the potential and accuracy of clustering-based redshift estimation, following the method proposed by M13. This technique allows us to infer redshift distributions from the spatial clustering of arbitrary sources with a set of reference objects for which redshift information is available. We applied it to the Main Spectroscopic galaxy sample from the Sloan Digital Sky Survey and, after homogenizing the spatial distribution of the reference population over the sky, we find that:
\begin{itemize}
\item We can characterize, with a high S/N, the redshift distributions of galaxy samples selected in narrow colour bins corresponding surface densities lower than $1$ source/deg$^2$. Ignoring the redshift evolution of the clustering amplitude of the sample, i.e. using $\d{\rm {\overline b_u}}/\d z=0$ in our formalism, the clustering-redshift technique provides us with an estimate of mean redshifts with an error $\delta z\sim0.02$.
\item Assuming that the galaxy bias evolves roughly linearly with redshift, i.e. $\d{\rm {\overline b_u}}/\d z=1$,  the clustering-redshift technique provides us with an estimate of mean redshifts with an error $\delta z\simeq0.001-0.01$ over the entire colour space of SDSS galaxies.
\item The characterization of redshift distributions as a function of galaxy colours provides us with a mapping which can be used to infer the redshift PDF of a single galaxy. This mapping is generic and can be used anywhere on the sky.
\item We find our clustering redshift estimates to provide more reliable results than the (KD-Tree) photometric redshifts \citep{csabai07} for the galaxy sample considered.
\end{itemize}
This analysis demonstrates, using real data, that clustering-based redshift inference provides us with a powerful data-driven technique to explore the redshift distribution of arbitrary datasets, without any prior knowledge on the spectral energy distribution of the sources.



\acknowledgments 

This work is supported by NASA grant 12-ADAP12-0270 and National Science Foundation grant AST-1313302. RS and SJS were supported by National Science Foundation Grant AST-1009514 and Department of Energy Grant DESC0009999.

Funding for the SDSS and SDSS-II has been provided by the Alfred P. Sloan Foundation, the Participating Institutions, the National Science Foundation, the U.S. Department of Energy, the National Aeronautics and Space Administration, the Japanese Monbukagakusho, the Max Planck Society, and the Higher Education Funding Council for England. The SDSS Web Site is http://www.sdss.org/.

The SDSS is managed by the Astrophysical Research Consortium for the Participating Institutions. The Participating Institutions are the American Museum of Natural History, Astrophysical Institute Potsdam, University of Basel, University of Cambridge, Case Western Reserve University, University of Chicago, Drexel University, Fermilab, the Institute for Advanced Study, the Japan Participation Group, Johns Hopkins University, the Joint Institute for Nuclear Astrophysics, the Kavli Institute for Particle Astrophysics and Cosmology, the Korean Scientist Group, the Chinese Academy of Sciences (LAMOST), Los Alamos National Laboratory, the Max-Planck-Institute for Astronomy (MPIA), the Max-Planck-Institute for Astrophysics (MPA), New Mexico State University, Ohio State University, University of Pittsburgh, University of Portsmouth, Princeton University, the United States Naval Observatory, and the University of Washington.

Facilities: \facility{The Sloan Digital Sky Survey}

\bibliographystyle{apj.bst}
\bibliography{rs-specz}

\appendix

\section{Reference Sample Clustering Amplitude}
\label{sec:refclustering}

\begin{figure}[h]
	\begin{center}
	\includegraphics[scale=0.5]{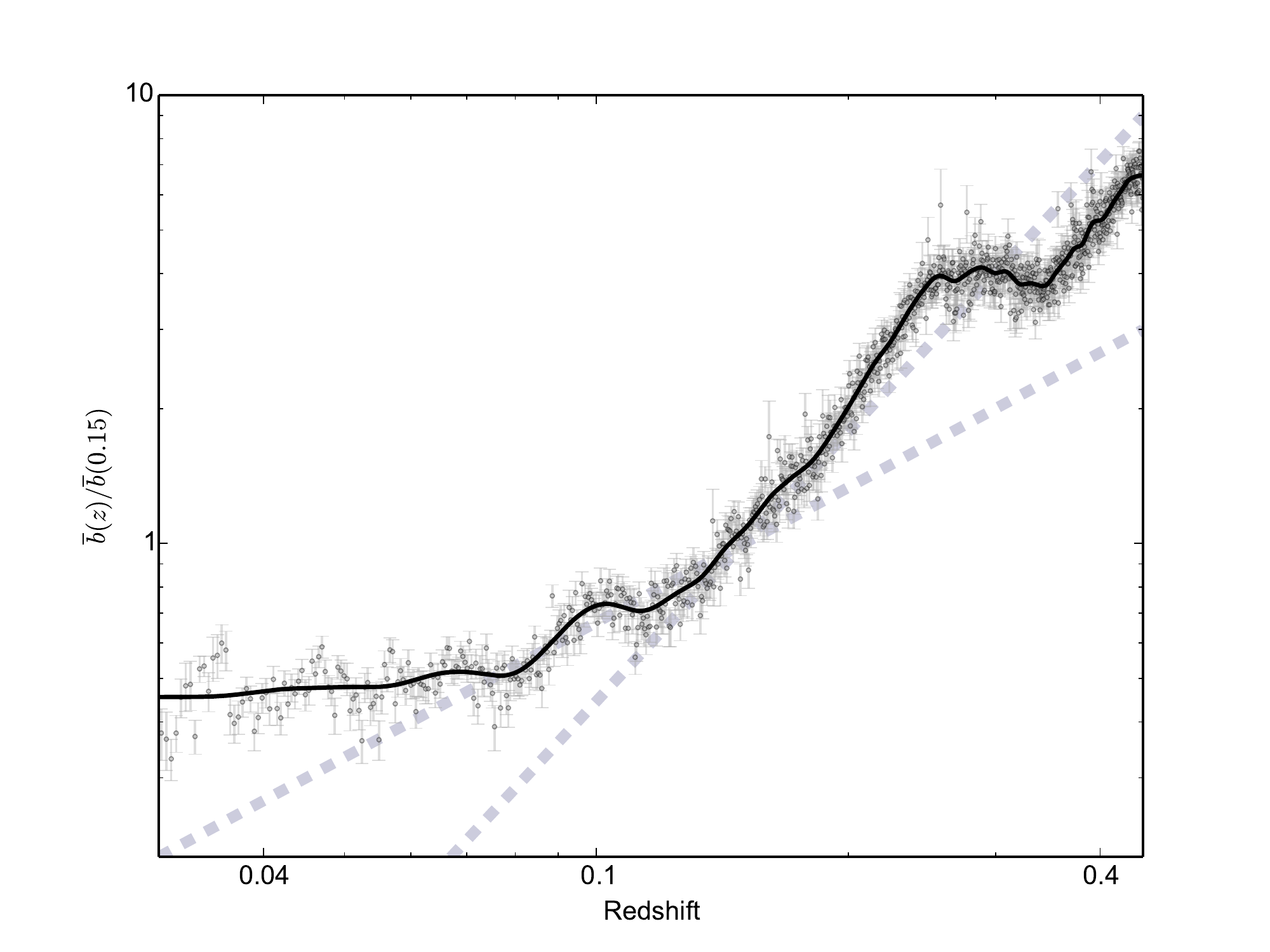}
	\end{center}

		\caption{The clustering amplitude redshift evolution as measured from the 
		SDSS Legacy spectroscopic galaxy sample, normalized to 1 at $z = 0.15$. 
		The solid line is the smoothed version of the clustering intensity evolution 
		curve used remove this factor from the clustering redshift distributions 
		in this paper. The dashed grey line corresponds to a 
		linearly evolving and quadratically evolving clustering amplitude 
		($d\overline{b}/dz = 1, 2$). 
		\label{fig:relbias}}
\end{figure}

The accuracy of our clustering-based redshift inference depends on the redshift dependence of the clustering amplitude of both the reference and the unknown samples. 
Here, we characterize the clustering amplitude $\d{\rm {\overline b_r}}/\d z$ of the reference sample we use in our analysis.
To do so, we measure the auto-correlation $w_{rr}(z)$ of the reference population as a function of redshift, considering the same range of scales and weighting as used in Equation \ref{eq:w_int} and estimate the clustering amplitude, normalized to an arbitrary redshift $z_0$, according to
\begin{equation}
\frac{{\rm {\overline b_r}}(z)}{{\rm {\overline b_r}}(z_0)} = \sqrt{\frac{\overline{w_{rr}}(z)}{\overline{w_{rr}}(z_0)}}\;.
\end{equation}
We note that this is different from the classical ``galaxy bias'' which is usually defined only on large-scales for which the galaxy and dark matter density fields are, on average, linearly related. Our bias definition includes contributions from small scales over which the galaxy and matter fields are non-linearly related.
We measure this quantity for our reference sample and we show its redshift dependence in Figure~\ref{fig:relbias}. We also show a smoothed version, for which we convolved the binned measurements with a Hann  filter of width $\Delta z = 0.02$, in Figure \ref{fig:relbias}. For reference, we show two lines indicating the expected evolution for $\d{\rm {\overline b_r}}/\d z=1$ and 2, normalized at $z=0.15$. We find this quantity to be weakly dependent on the maximum scale used in the  cross-correlation measurement, measured to an outer radius of 30 Mpc.  We note a plateau between $0.25 < z < 0.3$ where the SDSS spectroscopic galaxy selection changes from the main galaxy sample to the luminous red galaxy sample \citep{strauss02,eisenstein01}.


\section{Clustering Redshift Response Function}
\label{sec:response}

\begin{figure*}
\begin{center}
	\includegraphics[scale=0.59]{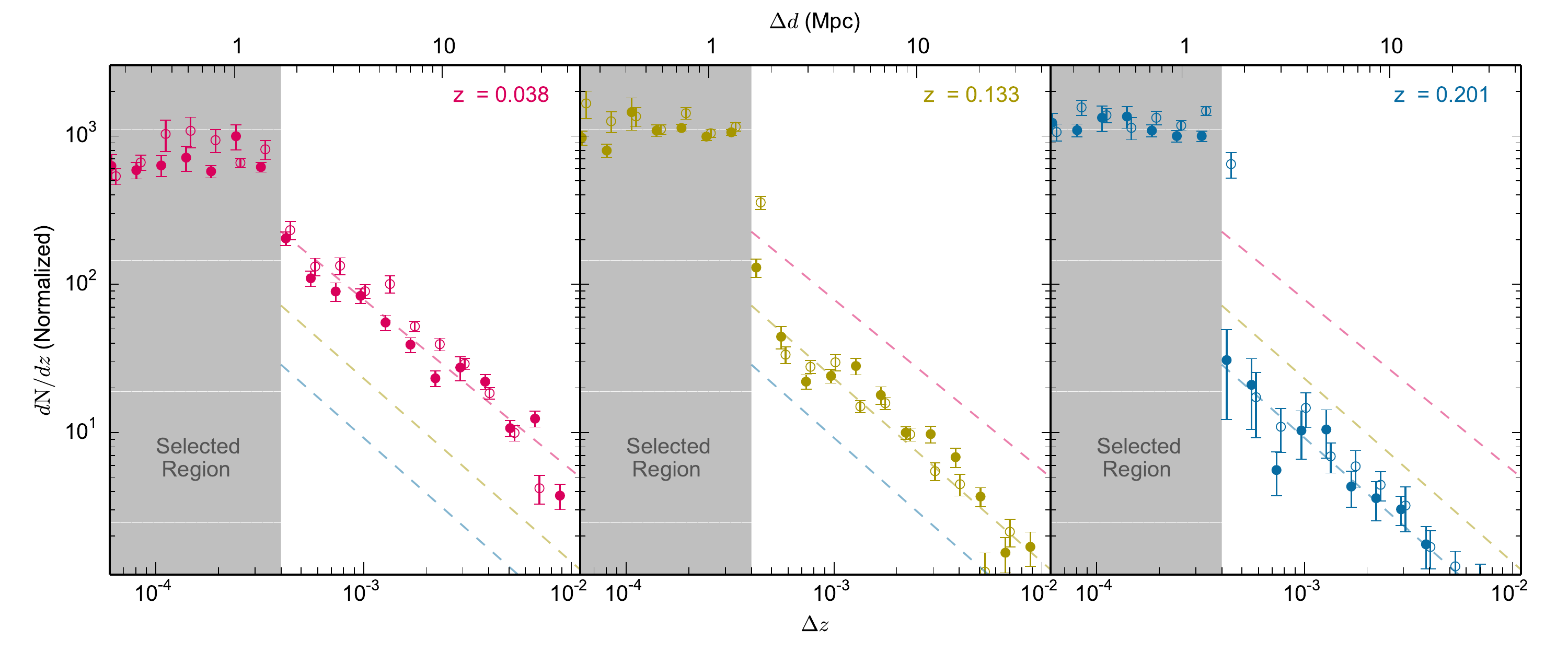}
\end{center}

	\caption{The clustering redshift distribution profiles as a function of
	distance from the central redshift for three different spectroscopically-
	selected samples. The selected redshift range is labeled in the top-right
	of each frame. The dashed line represents the power law fit to the
	profile between $3 \times 10^{-4} < \Delta z < 3 \times 10^{-3}$. The shaded
	regions indicates the true width of the selected distribution. In all cases,
	the tail of the distribution is consistent with the two-point correlation 
	function along the line-of-sight. This is consistent with the physical scales
	probed by these redshift scales (indicated on the top axis). 
	\label{fig:crdprofile}}

\end{figure*}

We investigate the response function of our $\d{\rm N}/\d z_{cl}$ estimator, corresponding the intrinsic uncertainty in the technique. In Figure~\ref{fig:crdprofile} we present the redshift profiles of the measured redshift density distribution for three different bins of reference redshift. The filled and open symbols show the redshift profiles on either side of the central redshift. The redshift range over which these reference samples are selected is shown by the grey region. Within this range we observe that our technique recovers a roughly flat distribution, as expected. Beyond this limit, our estimator is not consistent with zero (as ideally expected given the absence of reference galaxies in that range) but rather shows a power-law distribution with an index similar to $-1$ extending over the entire range of redshift intervals considered.  This is consistent with the expectation of clustering along the line of sight. This profile drops to below detectable levels beyond $\Delta z \sim 10^{-2}$ or $\Delta d \sim 30$ Mpc. The steep decline of the profile tail enables the clustering redshift technique to be sensitive to small variations in the redshift distribution at a level unreachable by current photometric redshift techniques.

This limit enables precisions previously only possible from spectroscopic redshifts. As has been noted in the text, the vast majority of correlation signal falls within the bounds of the selected sample; the correlation amplitude is a minimum of an order of magnitude greater within the selected sample bounds than in the tail of the response function.


\section{Scale Dependence of Clustering Redshifts}
\label{sec:scale}


\begin{figure*}
\begin{center}
	\includegraphics[scale=0.6]{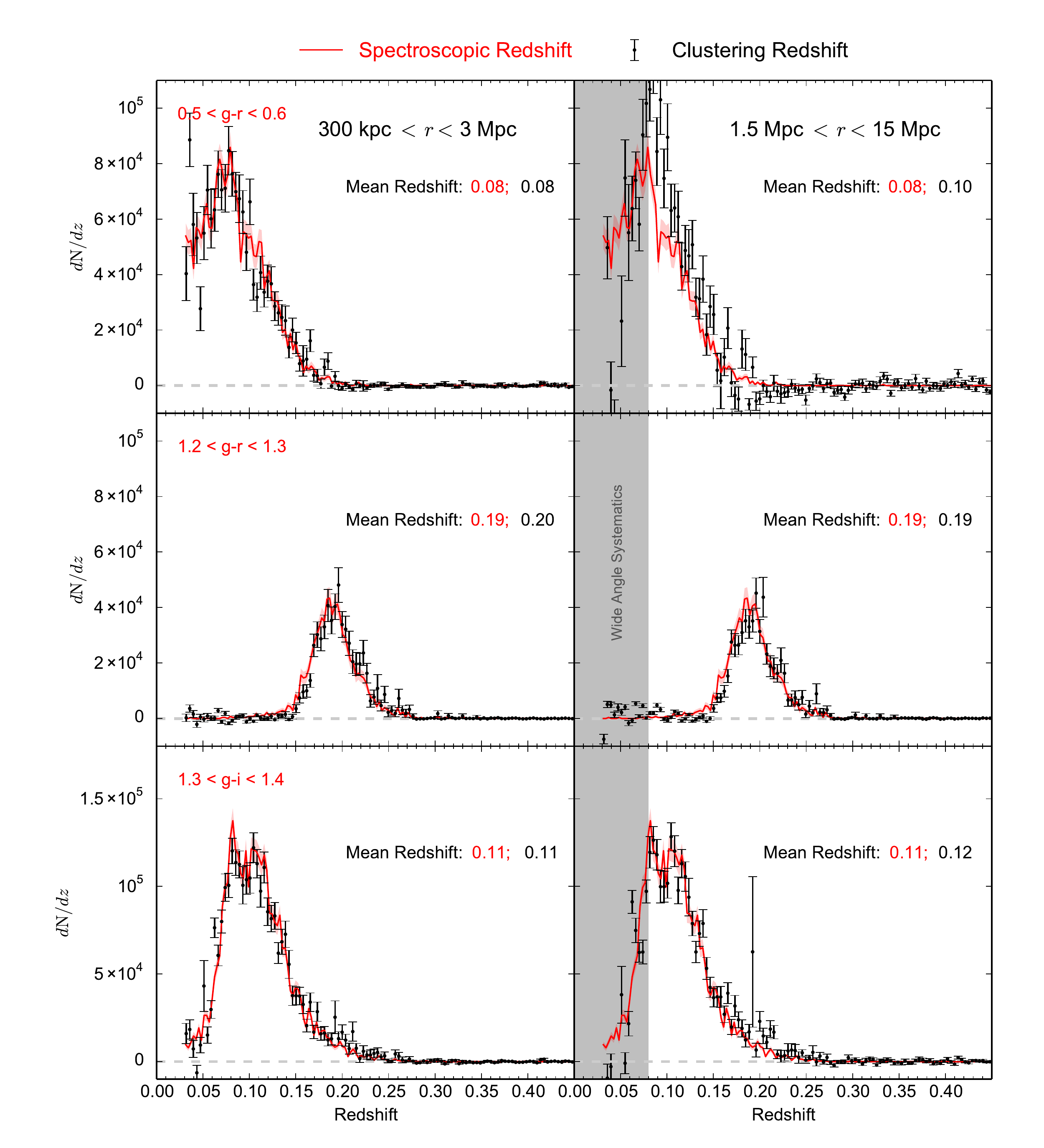}
\end{center}

	\caption{Comparison of the the clustering redshift distributions using the smaller cross-correlation annuli (300 kpc $< r <$ 3 Mpc; left column) and larger annuli (1.5 Mpc $< r <$ 15 Mpc; right column). Colour cuts and annotations are the same as in Figure \ref{fig:specialsamples_raw}. To minimize the effect of cosmic variance, the distributions are normalized to the signal above $z > 0.035$. The distributions are identical using either annuli sizes, demonstrating the scale independence of the technique and the applicability of clustering evolution measurements from larger scales. The cross-correlation measurements for the larger annuli have noise induced by the large angular scale used for measurements at low-$z$, which we indicate with the grey region; at higher redshift, the measurement is robust.
	\label{fig:scalecompare}}

\end{figure*}


Here we show that our clustering-based redshift inference technique does not strongly depend on the choice of scales used to measure the integrated correlation functions. The method has been shown to be robust to changes in scale through an analysis based on numerical simulations \citep{schmidt13} and we now demonstrate this property from data. We apply the technique to the colour cut samples from Section \ref{sec:results} using the original cross-correlation annulus (300 kpc $< r <$ 3 Mpc) and a much wider cross-correlation annulus (1.5 Mpc $< r <$ 15 Mpc). Ranging a decade of scale in both cases, we expect to extract a similar amount of clustering information from measurements in these two annuli given the slope of the matter correlation function. The results of the comparison are presented in Figure \ref{fig:scalecompare}. The measurements from the two different scales are nearly identical, despite covering substantially different areas around each reference source. The differences at low redshift ($z < 0.05$) come primarily from geometric effects of projecting large angles onto a spherical surface, the features present due to cosmic variance (i.e. voids), and the declining accuracy of measuring precise densities decreases as the solid angle increase (i.e. 15 Mpc is equivalent to 7\degr at $z = 0.03$). The discrepancies between the two scales disappear as redshift increases. The scale independence of this measurement illustrates the applicability of clustering redshifts including small scales where the galaxy and dark matter fields are not linearly related.

\end{document}